\documentclass[aps,prl,showpacs,notitlepage,twocolumn,superscriptaddress,nofootinbib,preprintnumbers]{revtex4-2}
\usepackage{bbm}
\usepackage{mathrsfs}

\usepackage{newtxtext}
\usepackage{epsfig}
\usepackage{csquotes}
\usepackage{soul,xcolor}
\usepackage{slashed}
\usepackage{graphicx}
\usepackage{amsfonts}
\usepackage{amsthm}
\usepackage[figuresright]{rotating}
\usepackage{amssymb}
\usepackage{amsmath}
\usepackage{dcolumn}
\usepackage{physics}
\usepackage{float}
\usepackage{bm}
\usepackage{physics}
\usepackage{verbatim}
\usepackage{braket}
\usepackage[normalem]{ulem}
\usepackage[ruled,vlined,linesnumbered]{algorithm2e}
\usepackage{setspace}
\usepackage{lipsum}
\usepackage[colorlinks,linkcolor=blue,anchorcolor=blue,citecolor=blue,urlcolor=blue]{hyperref}

\usepackage{CJK}

\usepackage{tikz,xcolor,hyperref}

\definecolor{lime}{HTML}{A6CE39}
\DeclareRobustCommand{\orcidicon}{
	\begin{tikzpicture}
	\draw[lime, fill=lime] (0,0) 
	circle [radius=0.16] 
	node[white] {{\fontfamily{qag}\selectfont \tiny ID}};
	\draw[white, fill=white] (-0.0625,0.095) 
	circle [radius=0.007];
	\end{tikzpicture}
	\hspace{-2mm}
}
\foreach \x in {A, ..., Z}{%
	\expandafter\xdef\csname orcid\x\endcsname{\noexpand\href{https://orcid.org/\csname orcidauthor\x\endcsname}{\noexpand\orcidicon}}
}
\foreach \x in {A, ..., Z}{%
	\expandafter\xdef\csname orcid\x\endcsname{\noexpand\href{https://orcid.org/\csname orcidauthor\x\endcsname}{\noexpand\orcidicon}}
}

\begin{document}

\title{Nuclear Mott transition with  neural quantum states
}
\author{Zi-Xiao Zhang\orcidA{}}
\affiliation{Key Laboratory of Nuclear Physics and Ion-beam Application (MOE), Institute of Modern Physics, Fudan University, Shanghai 200433, China}
\author{Xiao-Lu Qian\orcidB{}}
\affiliation{Department of Physics, Fudan University, Shanghai 200433, China}
\author{Wan-Bing He\orcidC{}}
\email[]{Contact author: hewanbing@fudan.edu.cn}
\affiliation{Key Laboratory of Nuclear Physics and Ion-beam Application (MOE), Institute of Modern Physics, Fudan University, Shanghai 200433, China}
\affiliation{Shanghai Research Center for Theoretical Nuclear Physics, NSFC and Fudan University, Shanghai 200438, China}
\author{Gerd R\"opke\orcidD{}}
\email[]{Contact author: gerd.roepke@uni-rostock.de}
\affiliation{Institut für Physik, Universität
Rostock, Rostock D-18051, Germany.}

\author{Kai-Jia Sun\orcidE{}}
\email[]{Contact author: kjsun@fudan.edu.cn}
\affiliation{Key Laboratory of Nuclear Physics and Ion-beam Application (MOE), Institute of Modern Physics, Fudan University, Shanghai 200433, China}
\affiliation{Shanghai Research Center for Theoretical Nuclear Physics, NSFC and Fudan University, Shanghai 200438, China}
\author{Yu-Gang Ma\orcidF{}}
\email[]{Contact author: mayugang@fudan.edu.cn}
\affiliation{Key Laboratory of Nuclear Physics and Ion-beam Application (MOE), Institute of Modern Physics, Fudan University, Shanghai 200433, China}
\affiliation{Shanghai Research Center for Theoretical Nuclear Physics, NSFC and Fudan University, Shanghai 200438, China}
\affiliation{School of Physics, East China Normal University, Shanghai 200062, China}

\begin{abstract} 
Bound clusters can dissolve in nuclear matter due to Pauli blocking, a phenomenon known as the Mott transition that is a key determinant of  the nuclear abundances in heavy-ion collisions.  Microscopic descriptions of this behavior require solving the in-medium Schrödinger equation. However, this remains a major computational challenge because the nuclear medium  breaks Galilean invariance and couples angular-momentum channels at finite center-of-mass momentum. The emergence of weakly bound states near dissociation further places stringent demands on conventional methods. Here, we address these challenges by introducing MottNet, a momentum-space neural quantum state framework based on self-attention that directly solves the $A$-body in-medium Schr\"{o}dinger equation. Benchmarked against exact diagonalization of the in-medium deuteron, MottNet reveals substantially weaker Mott suppression for $\alpha$ particles  in hot and dense nuclear matter than do conventional calculations based on combined approximations. At high baryon densities, the predicted $\alpha$-particle abundance is enhanced by orders of magnitude, even after continuum-edge subtraction. This finding offers a microscopic explanation for the unexpectedly large $\alpha$-particle fraction inferred from a combined analysis of FOPI heavy-ion collision data and transport model calculations.
 
\end{abstract}

\maketitle

\emph{Introduction.}---
The Mott transition~\cite{RevModPhys.40.677, RevModPhys.90.021001, Redmer2010}, the medium-induced dissolution of composite particles, is a ubiquitous phenomenon in quantum matter. It manifests in systems ranging from excitons in semiconductors~\cite{Li2021} and atoms in dense plasmas~\cite{Doppner2023} to light nuclei in nuclear matter~\cite{ROPKE1982536,PhysRevLett.80.3177,PhysRevC.81.015803,ONO2019139, PhysRevLett.108.062702,PhysRevLett.125.012701,PhysRevLett.108.172701,PhysRevLett.134.082304,ffxr-mmlg}. Across these 
regimes, as the environment grows denser, the sharp signatures of few-body bound states are progressively washed out. In nuclear matter, the Mott transition stems primarily from Pauli blocking by the surrounding Fermi sea, which determines whether few-nucleon correlations can persist as bound clusters~\cite{Oertel:2016bki}.
The formation and dissolution of these in-medium clusters have bearing on diverse topics, including the structure and decay of finite nuclei, heavy-ion dynamics, and the composition of  astrophysical matter~\cite{PhysRevLett.87.192501, PhysRevLett.110.262501,PhysRevLett.113.032506, Zhou2023, li2026minimalwignersu4interactionmicroscopic,Pons:1998mm,LIGOScientific:2017vwq,Most:2022wgo,PhysRevLett.111.132503,PhysRevLett.125.192502,REN2024138463, PhysRevLett.132.232502}.
 
A reliable description of these phenomena hinges on  a precise account of in-medium light cluster properties, which remains an open challenge. A recent few-body study has  addressed the  zero-temperature,
zero-momentum limit~\cite{ccbm-lk5v}. However, calculations of 
finite-momentum clusters are still beyond the reach of existing methods.   
The medium introduces a dependence on the center-of-mass (COM) momentum into the effective interaction, breaking the Galilean invariance and coupling different angular-momentum channels by reducing full rotational symmetry to axial symmetry. Consequently, conventional few-body methods, largely based on Jacobi-coordinate reductions and partial-wave decompositions, become unwieldy. Angle-average approximation is commonly used to decouple the angular-momentum channels~\cite{PhysRevC.60.034004, BEYER2000247,PhysRevC.63.034605,Beyer2004}.
However, $G$-matrix calculations have shown that retaining the full angular dependence of the two-body Pauli operator can produce non-negligible corrections to nuclear-matter saturation properties ~\cite{PhysRevC.59.2934, PhysRevC.62.014614}. Variational and perturbative treatments of medium effects have also been used to construct approximate dispersion relations for in-medium clusters~\cite{PhysRevC.81.015803,ROPKE201166, PhysRevC.79.014002,PhysRevC.92.054001}. The reliability of these approximations  has recently been called into question by the analysis of experimental heavy-ion data~\cite{ffxr-mmlg}, which points to a noticeably larger $\alpha$-cluster abundance than conventional predictions. 

Neural quantum states (NQS)~\cite{Lattice}, by representing the wavefunctions with neural networks,  have recently opened a new avenue for tackling this longstanding problem. NQS can offer compact yet expressive descriptions for quantum systems  without relying on a priori symmetry reductions of the configuration space. Furthermore, when optimized within variational Monte Carlo (VMC),  NQS bypasses the fermion sign problem associated with long-imaginary-time projection and can naturally accommodate the extended tails of weakly bound states. This is crucial for investigating the Mott transition, in which the
deeply bound and near-threshold
states must be described on the same footing. NQS have achieved remarkable
success in representing the wavefunctions of nuclei~\cite{KEEBLE2020135743, Lovato:2026erx,PhysRevLett.127.022502, PhysRevResearch.4.043178, Gnech2021NNNucleiA6, PhysRevC.107.034320, PhysRevLett.133.142501, Yang_2025, tlqz-nw28, ygkf-llyp, 42rh-6j8x,  Fore:2026spn, yang2026emergencehalo11linuclear}, hypernuclei~\cite{wmxg-cnrg, ZHANG2026140285, ZHANG2026140911, 3nrk-c5c7} and multiquark bound states~\cite{ckpr-s876}.
In these applications, the neural-network wavefunctions are predominantly presented in the coordinate space. 
For in-medium clusters, however, Pauli blocking depends on the momentum-space occupation probabilities of the surrounding nucleons, making momentum space the suitable representation for its accurate treatment.

In the present study, we develop a momentum-space NQS framework, termed MottNet, for a fully microscopic description of in-medium nuclei at arbitrary  temperature and COM momentum. 
MottNet directly learns momentum-space correlations through a self-attention mechanism.
This enables us to accurately solve the in-medium Schr\"odinger equation without relying on approximate treatments used in conventional approaches.  We find that the resulting $\alpha$-particle dispersion exhibits substantially weaker Mott suppression at high densities than conventional predictions~\cite{PhysRevC.81.015803}, enhancing the $\alpha$ abundance by several orders of magnitude.
This trend is consistent with that inferred from a combined analysis of heavy-ion collision data and transport model calculations~\cite{ffxr-mmlg}.

\emph{Method.}---
The properties of light nuclei  in  nuclear matter are governed by the in-medium Schr\"odinger equation. 
Starting from the finite-temperature Green's function formalism, we derive the in-medium kernel including both two- and three-nucleon forces
	\begin{align}
    \label{in-medium Hamiltonian}
    &\hat{K} = \sum_{i=1}^A \varepsilon_\tau(\hat{p}_i)
    + \sum^A_{i < j}\left[ 1 - \hat{f}_i - \hat{f}_j \right]\hat{v}_{ij}  \\
    &+\sum^A_{i < j < k, cyc}\left[ 1 - \hat{f}_i - \hat{f}_j - \hat{f}_k +\hat{f}_i\hat{f}_j+\hat{f}_i\hat{f}_k+\hat{f}_j\hat{f}_k\right]\hat{V}_{ijk}, \nonumber
	\end{align}
and define the in-medium Hamiltonian as its Hermitized part
    $\hat{H} = (\hat{K} + \hat{K}^\dagger) / 2$~\cite{PhysRevC.79.014002}.
In the above, $\varepsilon_\tau(\hat{p}_i)$ is the quasiparticle energy  relative to the continuum, $\hat{f}_i$ is the  momentum-space distribution of nucleons in the medium, 
$\hat{v}_{ij}$ and $\hat{V}_{ijk}$ are the two- and three-nucleon potentials. The derivation of Eq. (\ref{in-medium Hamiltonian})  can be found in the Supplemental Material~\cite{SM}.
When solving the in-medium Schr\"odinger equation, as in previous studies~\cite{PhysRevC.79.014002, PhysRevC.81.015803, PhysRevC.92.054001}, we use the uncorrelated-medium assumption,  whose validity is examined in Ref.~\cite{4py6-jyk8}. 
That is, the total nucleon distribution $\hat{f}_i$ is approximated by a Fermi-Dirac distribution, in which 
the chemical potential is determined by matching the baryon density $n_B$. 
As for the
underlying nuclear interaction, we adopt model “o” from Ref.~\cite{PhysRevC.103.054003}, derived
from leading-order pionless effective field theory. This model accurately reproduces the binding energy, radii, magnetic moment and photo-absorption cross section of $A\leq4$ atomic nuclei~\cite{PhysRevLett.133.142501, tlqz-nw28}.
 We also incorporate a static Coulomb force for protons (see Supplemental Material~\cite{SM}).

Motivated by the aforementioned considerations, we solve Eq.~(\ref{in-medium Hamiltonian}) with momentum-space NQS.
However, direct VMC calculations in momentum space have remained largely unexplored despite their potential advantages, owing to the need for both an expressive
 momentum-space ansatz and an efficient treatment of  nonlocal
momentum-space interactions.  Fortunately, deep learning provides a natural solution to the former, as neural network realizes a flexible mapping from many-body configurations to complex numbers, largely independent of the underlying representation. To this end, we introduce MottNet, which directly parameterizes the wavefunction in momentum space. The trial wavefunction $\Psi$ is constructed as follows:
\begin{align}
\langle\mathbf{p}|\Psi\rangle = e^{J(\mathbf{p})}\sum_{k=1}^K \det [\phi_i^{(k)}(\mathbf{p}_j; \{\mathbf{p}_{/j}\})].
\label{NQS}
\end{align}
We here use the abbreviation $\mathbf{p}_j=\{\bm{p}_j, s_j, t_j\}$ to denote the momentum-spin-isospin eigenstate for the $j$-th nucleon and $\mathbf{p}$ the complete nuclear configuration. The notation $\{\mathbf{p}_{/j}\}$ indicates that the $\phi^{(k)}_i$ is invariant under permutations of the nucleon indices except $\mathbf{p}_j$. A Jastrow factor $J(\mathbf{p})$ is used to modulate the overall amplitude, constructed with a permutation-invariant neural network. Since momentum-space correlations are not naturally characterized by pairwise distances, inspired by the condensed matter community~\cite{vonglehn2023selfattentionansatzabinitioquantum, Li2024, Li2024b, PhysRevLett.134.176503}, we employ a self-attention mechanism to realize the many-body backflow transformation~\cite{PhysRevLett.122.226401} in Eq. (\ref{NQS}), which allows correlations to
be built in without manually crafted two-body features. Details of MottNet are provided in the Supplemental Material~\cite{SM}.

With the ansatz constructed, we next turn to the nonlocality of momentum-space
interactions. The treatment of nonlocal potentials is a common challenge in NQS calculations. In chemistry, the pseudopotentials lead to additional angular integrals in the local energy~\cite{PhysRevResearch.4.013021}, which can be evaluated by spherical quadrature. In lattice systems, the analogous contributions appear as sums over sites connected by the Hamiltonian~\cite{Choo2020}. 
To illustrate the basic design of our computational strategy, we use a spin-isospin independent 2$N$ force as an example
\begin{align}
\langle\bm{p}_i,\bm{p}_j|\hat{v}_{ij}|\bm{p}'_i,\bm{p}'_j\rangle =  C_2R(\bm{p}_i, \bm{p}_j, \bm{p}'_i, \bm{p}'_j)\delta(\bm{p}_i+\bm{p}_j-\bm{p}_i'-\bm{p}'_j),
\label{matrix element}
\end{align} 
 in which $C_2$ is the coupling constant and $R(\bm{p}_i, \bm{p}_j, \bm{p}'_i, \bm{p}'_j)$
 the regulator. Within the VMC formalism, we construct the energy expectation by averaging the local energy over a large batch of samples drawn from the distribution $|\Psi|^2$
 \begin{align}
\frac{\langle\Psi|\hat{H}|\Psi\rangle}{\langle\Psi|\Psi\rangle} = \mathbb{E}_{|\Psi|^2}\!\left[ \frac{\langle \mathbf{p}|\hat{H}|\Psi\rangle}{\langle \mathbf{p}|\Psi\rangle} \right].
 \end{align}
The kinetic operator is diagonal in momentum space, leaving the construction of the local interaction term as the main challenge
\begin{align}
    \langle\mathbf{p}|\hat{v}_{ij}&|\Psi\rangle = \int d\bm{p}'_i\bm{p}'_j\langle\mathbf{p}|\hat{v}_{ij}|\mathbf{p}_{i'j'}\rangle\langle\mathbf{p}_{i'j'}|\Psi\rangle \nonumber\\
    &=C_2\int d\bm{p}_i' R(\bm{p}_i, \bm{p}_j, \bm{p}'_i, \bm{K}_{ij} - \bm{p}'_i)\langle\mathbf{p}_{i'j'}|\Psi\rangle,
    \label{matrix element full}
\end{align}
where $\mathbf{p}_{i'j'}$ denotes $\{\bm{p}_1, ..,\bm{p}_{i}', .. ,\bm{p}_{j}', ..,\bm{p}_A\}$ with spin-isospin unchanged  and $\bm{K}_{ij} = \bm{p}_i + \bm{p}_j$. Eq.~(\ref{matrix element full}) involves a continuous multidimensional integral over intermediate momenta, which we evaluate using importance-sampling Monte Carlo. Specifically, we generate an ensemble of momenta sampled according to the regulator $R(\bm{p}_i, \bm{p}_j, \bm{p}'_i, \bm{K}_{ij} - \bm{p}'_i)$, which can naturally suppress the contribution from the high-momentum region to achieve efficient Monte Carlo.

\emph{Results.}---
We first validate MottNet by comparing the computed ground-state energies and charge radii of free-space \(^2\)H, \(^3\)H, \(^3\)He and \(^4\)He with coordinate-space NQS calculations. 
The results are reported in the Supplemental Material~\cite{SM}. Both methods show excellent agreement within statistical uncertainties and reproduce the experimental ground-state properties of light nuclei.
Non-perturbative benchmarks for  clusters at finite
COM momentum are generally unavailable beyond the two-body level.
We therefore examine the in-medium deuteron, for which exact lattice diagonalization remains feasible. MottNet accurately reproduces
the diagonalization results, including near the continuum threshold (see Supplemental Material~\cite{SM}).

 \begin{figure}[t]    \includegraphics[width=0.49\textwidth]{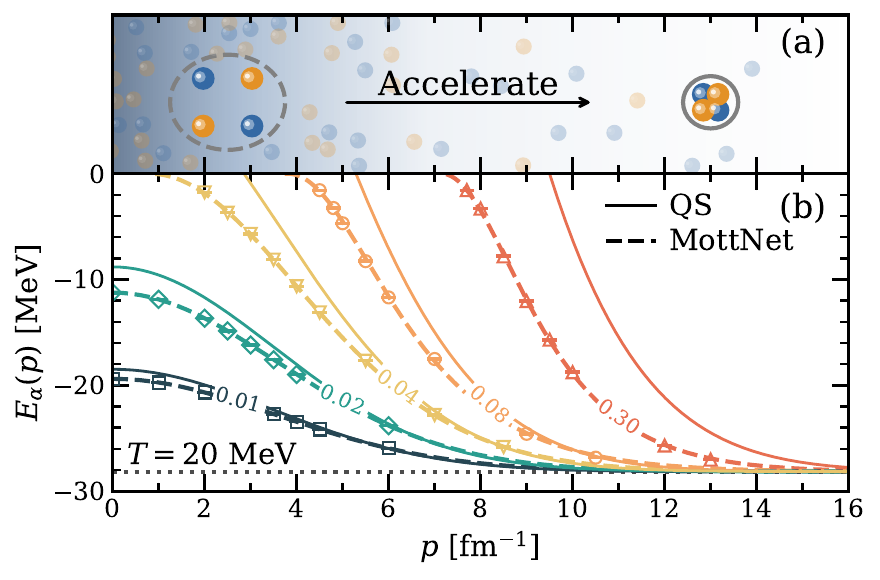}
    \caption{(a) Schematic illustration of an $\alpha$ cluster moving in a background consisting of uncorrelated nucleon gas. (b) In-medium \(\alpha\)-particle energy \(E_{\alpha}(p)\) in symmetric nuclear matter at \(T = 20\) MeV as a function of the COM momentum \(p\). The numbers on the lines are baryon densities in fm$^{-3}$.
Symbols denote the MottNet calculations, dashed lines are fits using
Eq.~(\ref{ansatz}), and solid lines show the corresponding QS
parametrizations~\cite{PhysRevC.81.015803}. The horizontal dotted line indicates the vacuum
$\alpha$-particle energy. \vspace{-18px}}
    \label{fig:dispersion}
\end{figure}
With these benchmarks established, we now turn to the Mott transition of light clusters. We use $\nu$ to denote the particle
species, with $p$, $n$, $d$, $t$, $h$, and $\alpha$ referring to protons,
neutrons, deuterons, tritons, helions, and $\alpha$
particles, respectively.
Fig.~\ref{fig:dispersion} (a) sketches how Pauli blocking changes from low to high COM momentum. As an in-medium $\alpha$ cluster is boosted to larger COM momenta, its constituent nucleons probe momentum
regions with a
more dilute background nucleon distribution, which weakens the
Pauli blocking and  enhances the cluster
binding. 
The composition and thermodynamic properties of  nuclear matter are strongly controlled by the in-medium dispersion relations $\varepsilon_\nu(\bm{p})$. 
Fig.~\ref{fig:dispersion} (b) shows the in-medium $\alpha$-cluster energy relative to the continuum, $E_\alpha(\bm{p})$, at $T=20$ MeV. MottNet yields cluster energies at discrete COM momenta. We fit these energies with the following piecewise ansatz to obtain a continuous representation
\begin{align}
    E_\nu(\bm{p}) &= 
    \begin{cases}
        [E_\nu(0)-E_\nu(\infty)]\frac{e^{-ap^2}}{1+bp^2} + E_\nu(\infty), & P^{\text{Mott}}_\nu=0, \\[1ex]
        -E_\nu(\infty) \frac{e^{-a(p-P^{\text{Mott}}_\nu)^2}}{1+b(p-P^{\text{Mott}}_\nu)^2} + E_\nu(\infty), & P^{\text{Mott}}_\nu>0,
    \end{cases}\nonumber\\
    \varepsilon_\nu(\bm{p}) &= \frac{p^2}{2A_\nu M} + E_\nu(\bm{p}),
    \label{ansatz}
\end{align}
in which $a, b, P_\nu^{\text{Mott}}$ are the fitting parameters and provided in the Supplemental Material~\cite{SM}. Inspired by Refs. \cite{Kim2024, PhysRevLett.133.142501}, we introduce a \enquote{transfer learning} strategy for an effective exploration near the Mott boundary. Specifically, the optimized wavefunction at a neighboring density or COM momentum is used to initialize a subsequent calculation.

For all densities, at small COM
momenta, the cluster binding is strongly suppressed by the medium and becomes unbound for $n_B \geq 0.04$ fm$^{-3}$.  As Pauli blocking weakens at large COM momenta, the in-medium $\alpha$ cluster progressively recovers its vacuum character.
The MottNet dispersion relation is compared with conventional parametrizations widely employed in quantum statistical (QS) approaches to hot nuclear matter~\cite{PhysRevC.81.015803,ROPKE201166, PhysRevC.79.014002,PhysRevC.92.054001}. While QS parametrizations capture the qualitative momentum dependence, noticeable deviations from the MottNet results emerge near the Mott boundary. Such discrepancies likely reflect the cumulative approximations entering the conventional parametrizations, which become increasingly consequential at high densities.  This can be seen from the excellent agreement at 
$n_B=0.01~\mathrm{fm}^{-3}$, while sizable differences emerge at 
$n_B=0.30~\mathrm{fm}^{-3}$, where strong medium effects require direct
microscopic treatments.
  \begin{figure}[tb]
\includegraphics[width=\columnwidth]{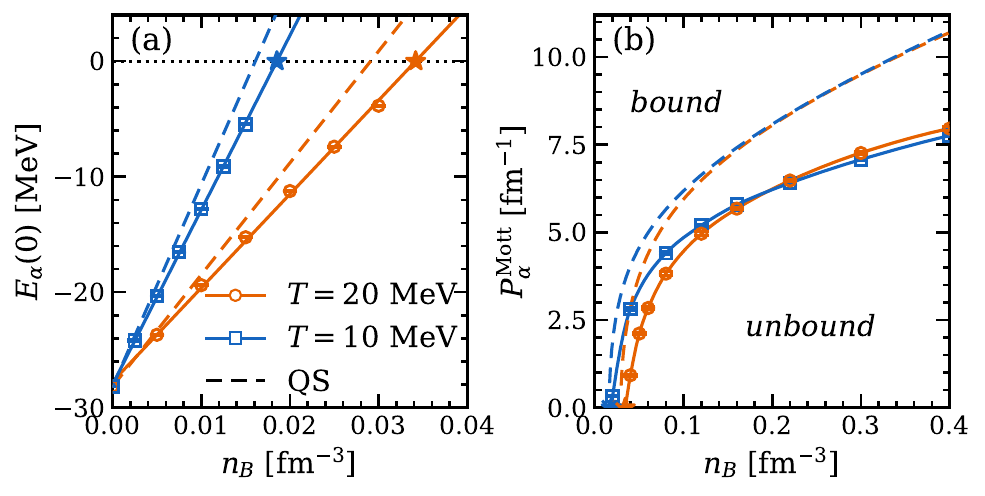}
    \caption{Mott transition of the $\alpha$ particle in symmetric nuclear matter. (a) In-medium energy of an $\alpha$ particle at rest, $E_\alpha(0)$, as a function of baryon density for  $T=10$ MeV (blue squares) and 20 MeV (orange circles), compared with the corresponding QS
parametrizations~\cite{PhysRevC.81.015803}. Solid lines show linear extrapolations of the calculated energies to $E_\alpha(0)=0$, yielding the corresponding Mott densities (star symbols).  
    (b) Density dependence of the Mott momentum of the $\alpha$ particle in symmetric nuclear matter. Solid lines are interpolations. The stars mark the extrapolated Mott densities in the leftmost panel.  Error bars propagate from the fitting uncertainty for the parameters in Eq. (\ref{ansatz}). The uncertainties are smaller than the line width and the symbol size.} 
    \label{fig:pmott}
\end{figure}

MottNet calculations provide access to the Mott boundaries of in-medium clusters. The Mott density $n_\nu^{\rm Mott}$ is defined by the condition $E_\nu(0)=0$, i.e., the density at which a cluster at rest becomes unbound.
However, pinpointing these thresholds is challenging because Monte Carlo approaches rely on normalizable probability distributions and thus are not directly suited for unbound states.
We therefore determine $n_\nu^{\rm Mott}$ by linearly extrapolating the calculated static energies to $E_\nu(0)=0$, and infer $P_\nu^{\rm Mott}$ by fitting the finite-momentum energies with Eq.~(\ref{ansatz}).
The corresponding uncertainties are obtained by propagating the errors of the fitted parameters. 

Fig.~\ref{fig:pmott} (a) shows the in-medium static energy $E_\alpha(0)$ as a
function of density at $T=10$ and $20$ MeV. As the density increases, binding is quenched by Pauli blocking and vanishes at $n_\nu^{\rm Mott}$. The MottNet energies  exhibit an approximately linear dependence on density, which  is consistent with the QS parametrizations.
  The  Mott density increases with temperature, which can be understood from the thermal smearing of the Fermi surface that weakens the Pauli blocking. 
  The determined $n_\alpha^{\text{Mott}}$ serve as the transition points for the $P_\alpha^{\rm Mott}$, shown in Fig.~\ref{fig:pmott} (b).
The MottNet and QS predictions for 
$P_\alpha^{\rm Mott}$ agree in the dilute regime but separate rapidly 
as $n_B$ increases. At $n_B=0.30~{\rm fm}^{-3}$, the MottNet calculation gives a
$P_\alpha^{\rm Mott}$ more than $2~{\rm fm}^{-1}$ lower than that from the QS
parametrization, which indicates a much 
weaker Mott effect in hot and dense nuclear matter. This trend echoes a recent analysis of the FOPI light-nuclei yields in central Au+Au heavy-ion collisions at 0.25$A$-0.6$A$ GeV~\cite{ffxr-mmlg}.
\begin{figure}[tb]
    \includegraphics[width=0.49\textwidth]{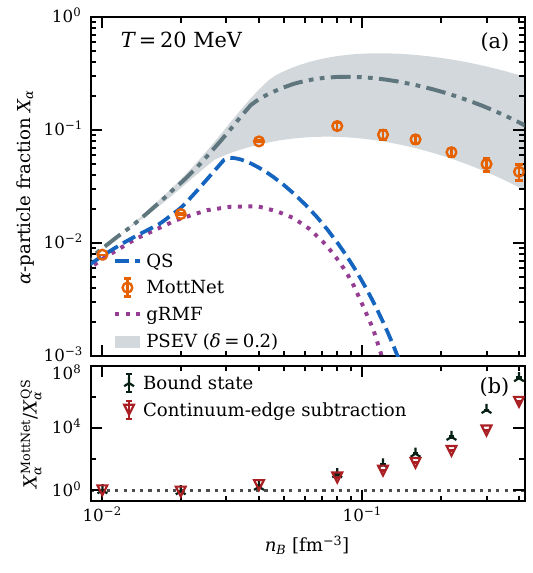}
    \caption{$\alpha$-particle abundance in nuclear matter at $T=20$ MeV.
(a) Mass fraction $X_\alpha$ as a function of baryon density, calculated from the bound-state population defined in Eq. (\ref{distribution}).
MottNet results are compared with the QS parametrization and gRMF calculations~\cite{PhysRevC.81.015803},
as well as the PSEV result inferred  from FOPI light-nuclei
yields~\cite{ffxr-mmlg}. The PSEV result is for asymmetric nuclear matter with $\delta=0.2$, and the shaded band indicates its 90\% confidence interval. Error bars show uncertainties propagated from the fitted parameters in Eq. (\ref{ansatz}).
(b) $X_\alpha^{\rm MottNet}/X_\alpha^{\rm QS}$ evaluated
with the bound-state population and with the continuum-edge subtraction.
The horizontal dotted line marks unity.}
    \label{fig:fraction}
\end{figure}
Using the
phase-space excluded-volume (PSEV) prescription to connect transport
calculations with equilibrium nuclear matter~\cite{REISDORF2007459,PhysRevC.108.L031601, Sun2024}, the inferred
$\alpha$-particle fraction reaches $X_\alpha\sim0.2$ in the dense freeze-out
region, well above conventional predictions such as generalized relativistic mean-field (gRMF).
While the gRMF discrepancy has been attributed to its neglect of momentum-dependent Pauli blocking, the disagreement with the more refined QS model~\cite{PhysRevLett.108.172701}, however, remains unexplained, as QS already accounts for the momentum dependence of Pauli blocking.

In Fig. \ref{fig:fraction} (a), we compare the above results on $\alpha$-particle fraction in $T=20$ MeV nuclear matter. Following the PSEV analysis of
Ref.~\cite{ffxr-mmlg}, we first use the bound-state population for the momentum distribution of light clusters
\begin{align}
    f_{\nu}(\bm{p}) = \frac{g_\nu H(|\bm{p}| - P^{\text{Mott}}_\nu)}{\exp (\beta[\varepsilon_\nu(\bm{p})-\mu_\nu])\pm1},
    \label{distribution}
\end{align}
where the plus [minus] sign is for fermions [bosons], $\beta$ the inverse temperature, $\mu_\nu$ the chemical potential, $g_\nu$ the degeneracy and $H$  the Heaviside step function. Under the chemical equilibrium condition, the chemical potential of a cluster is $\mu_\nu = Z_\nu \mu_p + N_\nu \mu_n$, where $Z_\nu$ and $N_\nu$ are the numbers of protons and neutrons, respectively.  Substituting either the MottNet or the QS dispersion into $\varepsilon_\nu(\bm{p})$, the chemical potential can be determined by matching the total proton and neutron densities. The mass fraction for cluster $\nu$ reads $X_\nu = n_\nu/n_B$, with $n_\nu=\int \frac{d^3 \bm{p}}{(2\pi)^3} A_\nu f_\nu(\bm{p})$. At low densities,  $X_\alpha$ from MottNet, QS and gRMF are nearly identical. This agreement aligns with previous analyses of NIMROD data at
$n_B=0.002$--$0.03~{\rm fm}^{-3}$, where QS calculations provided one of the
closest descriptions of the extracted $\alpha$ equilibrium constants~\cite{PhysRevLett.108.172701}.
With increasing $n_B$, the QS parametrization suppresses $X_\alpha$ rapidly, whereas the MottNet dispersion gives a much larger $\alpha$ abundance and moves toward the PSEV estimate constrained by FOPI light-nuclei yields.  The PSEV curve remains systematically above MottNet, even for $\delta=0.2$ nuclear
matter. This can be traced to the use of the free-space cluster energies in the PSEV framework~\cite{ropke2026clusterednaturehotdense}. Although the empirical calibration of the phase-space cutoff can partially compensate for this omission, a microscopic treatment requires explicit in-medium dispersion relations.

In the QS framework~\cite{PhysRevC.81.015803}, the cluster distribution in Eq. (\ref{distribution}) is supplemented by a
continuum-edge subtraction motivated by the Levinson theorem. To separate the dispersion effect from the
treatment of continuum correlations, Fig.~\ref{fig:fraction} (b) shows the
enhancement factor
$X_\alpha^{\rm MottNet}/X_\alpha^{\rm QS}$ evaluated with both
quasiparticle prescriptions. In both cases, the MottNet dispersion enhances
the $\alpha$ abundance by orders of magnitude around
$n_B=0.3~{\rm fm}^{-3}$. Since MottNet directly solves the in-medium Hamiltonian, this
finding indicates that conventional models substantially underestimate
finite-momentum $\alpha$ correlations in hot and dense nuclear matter. The
persistence of the enhancement after the continuum-edge subtraction shows that this
conclusion is not an artifact of the continuum treatment.  

\emph{Summary.}---
We have developed MottNet, a momentum-space NQS framework based on self-attention, to rigorously solve the in-medium $A-$body Schr\"{o}dinger equation that is essential for understanding the Mott transition of light nuclei in a nuclear medium. We benchmark MottNet against coordinate-space NQS calculations in the free space and exact diagonalization of the in-medium deuteron, and apply it to light clusters with $A\leq4$.
We find substantially weaker Mott suppression of $\alpha$ particles than conventional calculations  based on combined approximations including angle-average, perturbative treatment, and variational approximations~\cite{PhysRevC.81.015803}. 
This leads to an abundance enhanced by several orders of magnitude even after continuum-edge subtraction, offering a microscopic foundation for the unexpectedly large $\alpha$-particle fraction recently inferred from 
heavy-ion collision data~\cite{ffxr-mmlg}. These findings suggest that conventional approaches may substantially underestimate $\alpha$-particle survival in hot and dense nuclear matter, with important implications for the nuclear equation of state.

This work highlights the potential of neural networks for tackling quantum systems where reduced symmetry and weak binding challenge conventional methods.
Although the presented calculations employ only leading-order interactions, the regulator-guided importance sampling strategy introduced here provides a promising route for future calculations with realistic nonlocal chiral nuclear interactions. More broadly, the MottNet framework may be extended to study the Mott transition in diverse quantum systems, including dense plasmas and semiconductor systems, where medium-induced modifications of bound states play an important role.

\emph{Acknowledgements}
---The authors thank Jane Kim, Bing-Nan Lu, Zheng-Xue Ren, Stefan Typel, Rui Wang, Yi-Long Yang and Peng-Wei Zhao for helpful discussions. This work was supported in part by the National Key Research and Development Project of China under Grants No. 2023YFA1606601 and 2024YFA1612500; the National Natural Science Foundation of China under Contracts No. 12688101, No. 12547102, No. 124B100007, No. 12475187, No. 12422509, No. 12375121; the STcSm under Grant No. 23590780100 and 23JC1400200. The computations in this research were performed using the CFFF platform of Fudan University.

\emph{Data availability}
---The numerical data supporting the findings of this article are available from the corresponding authors upon reasonable request.

\bibliography{ref_main}

\bibliographystyle{apsrev4-1}

\end{document}